\documentclass[12pt,a4paper]{article}
\usepackage[margin= 1 in]{geometry}
\usepackage{graphicx}
\usepackage{latexsym}
\usepackage{amsmath}
\usepackage{scalerel}
\usepackage[utf8]{inputenc}
\usepackage[T1]{fontenc}
\usepackage{tcolorbox}

\title{Constraining Brans-Dicke Parameter Using Gravitational Radiation}
\author{Paritosh Verma \\ Narodowe Centrum Bada$\acute{n}$  J\c{a}drowych \\ Andrzeja Sołtana 7, 05-400 Otwock, Polska \\ Paritosh.Verma@ncbj.gov.pl}
\date{}

\begin{document}

\maketitle

\section{Abstract}

This article presents a theoretical limit on the Brans-Dicke (BD) parameter using a neutron star (NS) revolving around a galactic central engine.The assumption that the orbital radius is large enough allows  to ignore strong curvature due to the central supermassive black hole (SMBH) and treat the problem with Newtonian dynamics. First, gravitational radiation polarizations are calculated in the BD theory. The scalar polarization, dominated by the dipole radiation, is then used to put a constraint on the BD parameter.The calculations are relatively simple and this makes it accessible to sophomore students.


\section{Introduction }

In 1961, Robert H. Dicke and Carl H. Brans proposed a scalar-tensor theory, known as Brans-Dicke (BD) theory \cite{Brans-Dicke}, to describe gravitation by incorporating Mach's principle. The foundation of this theory was built on the previous work of Pascual Jordan \cite{Jordan}  as well as Markus Fierz \cite{Fierz}, and sometimes it is also referred to as Jordan-Fierz-Brans-Dicke theory. There are three polarization states of gravitational radiation in BD theory: two tensor states (similar to general relativity GR) and one scalar polarization dominated by the dipole radiation \cite{Isi}.

The field equations of BD theory contain a parameter called the Brans-Dicke coupling constant $\omega_{BD}$.  The BD parameter  is obtained through experiments, and the Cassini \cite{Bertotti-Tartora} experiment in 2003 has imposed the constraint $ \omega_{BD} > 40,000 $. To simplify our further calculations, we define a parameter $\zeta$ 

 \begin{equation}\label{eq:0.0}
\zeta \equiv \frac{1}{2 \omega_{BD} + 4}
 \end{equation} 

and  the gravitational coupling depends through the relation $G (1 - \zeta)$, where $G$ is the gravitational constant.  

Since the last hundred years, there have been many attempts to modify or test GR  \cite{GR test GWTC3}, \cite{GR-test-GW170817}, \cite{Bertolami}, \cite{JBD-cosmo}, \cite{Sola}, \cite{BD-LVK} and it has successfully passed most of them. There are different regimes to test theories of gravity \cite{Wex}: Quasi-stationary weak-field regime (G1), Quasi-stationary strong-field regime (G2), Highly-dynamical strong-field regime (G3) and Radiation regime (GW). In this paper, we use the radiation regime to impose a constraint on the BD theory whereas the Cassini mission (mentioned above) tested gravity in the G1 regime. 

\begin{figure}[!t]
\caption{A neutron star (purple circle) revolves around a SMBH (black circle) at a very large distance and the path between points A and B can be approximated as motion in a straight line. The angle subtended from the center as NS moves from A to B is $\theta$.}\label{SMBH}

\centering
\includegraphics[width=6.5cm, height=6.5cm]{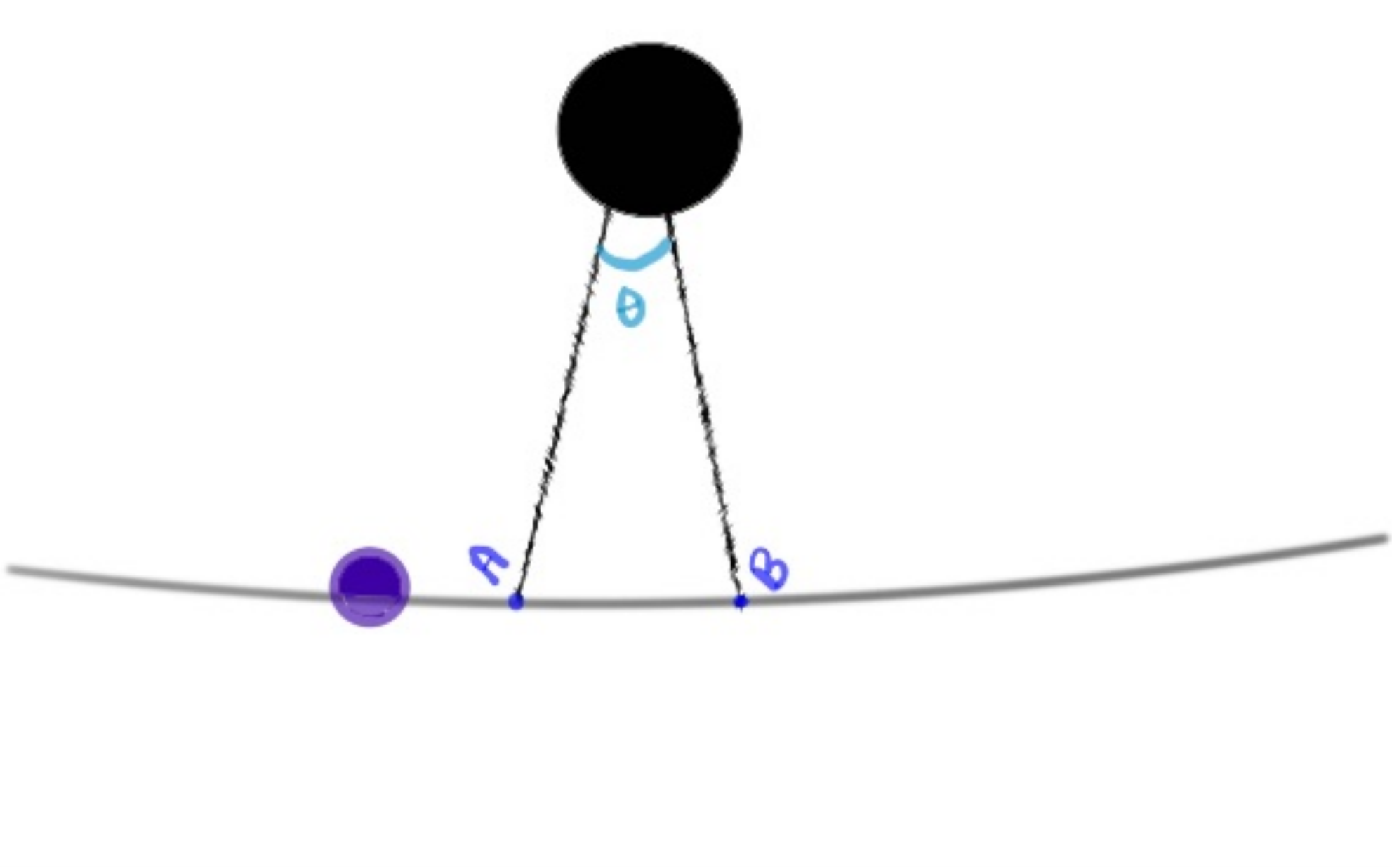}
\end{figure}

We consider a simple case when a NS is revolving around a SMBH at a large distance with a speed v much less than the speed of light c. At such a distance, the curvature is small and we can treat the problem using Newtonian dynamics. Since the NS is too far away from the central engine, we can safely model it with a Dirac delta function in comparison to the size of a SMBH. For a time $t'$ much less than  the orbital period $T$, we can safely approximate the path of the motion to be linear as shown in the Figure \ref{SMBH}.

\section{Polarizations}

When the NS moves in its orbit, its acceleration can be decomposed into two components. The radial acceleration $a_r$ is due to the centripetal force whereas the tangential component $a_t$ is due the radiation reaction which takes away the kinetic energy. We ignore the contribution due to other celestial bodies in the vicinity that can accelerate or decelerate our target. 

For a short distance in orbit, the position of the NS is given by

 \begin{equation}\label{eq:1}
x_p(t') = ut' + \frac{1}{2} a_t t'^2
 \end{equation} 

where u is the initial velocity and $ a_t \in R^- $. Since, gravitational radiation is feeble, we can safely assume $a_t$ to be a constant for a short period of time.  

To study the dynamics, we need to calculate the quadrupole moment and dipole moments of the NS. To compute these moments, we fix an inertial frame whose x-axis points along the direction of the motion of NS. 

The symmetric trace free quadrupole moment tensor is given by

 \begin{equation}\label{eq:2}
Q^{ij} = \int \left[ x^i x^j - \frac{1}{3} r^2 \delta_{ij}    \right] dm
 \end{equation} 

where $r^2 = x^2 + y^2 + z^2$ and dm is the infinitesimal mass element given by

 \begin{equation}\label{eq:3}
dm = \lambda dx
 \end{equation} 

$\lambda$ is the linear mass density given by $\lambda = m \delta \left( x - x_p(t') \right)$ and $m$ is the mass of NS. The $Q^{xx}$ component in the inertial frame is

 \begin{equation}\label{eq:4}
Q^{xx}(t') = m \int \delta \left( x - x_p(t') \right) \left[ x^2 - \frac{1}{3} \left( x^2 + 0 + 0  \right) \delta_{xx}    \right] dx
 \end{equation} 

or,

 \begin{equation}\label{eq:5}
Q^{xx}(t') = \frac{2}{3} m x_p^2(t')
 \end{equation} 

The $Q^{yy}$ component in the inertial frame is

 \begin{equation}\label{eq:6}
Q^{yy}(t') = m \int \delta \left( x - x_p(t') \right) \left[ 0 - \frac{1}{3} \left( x^2 + 0 + 0  \right) \delta_{yy}    \right] dx
 \end{equation} 

or,

 \begin{equation}\label{eq:7}
Q^{yy}(t') = - \frac{1}{3} m x_p^2(t')
 \end{equation} 

Similarly, we can compute other components and obtain

     \begin{equation}
 \label{eq:8}
Q^{ij} (t') =  Q_{0} (t') \begin{bmatrix} 2  & 0 & 0 \\ 0 & - 1  & 0 \\ 0 & 0 & - 1  \end{bmatrix}
 \end{equation} 

where $Q_{0} (t')$ is defined as

\begin{equation}
 \label{eq:8.1}
  Q_{0}(t') \equiv    \frac{1}{3} m x_p(t')^2 
 \end{equation}

The x-component of the dipole moment in the inertial frame is given by

     \begin{equation}
 \label{eq:9}
d(t') = m x_p (t')
 \end{equation} 
 
The total dipole moment in the vector form is 

    \begin{equation} 
 \label{eq:10}
D(t')  =  \begin{bmatrix} d(t')   \\ 0  \\ 0  \end{bmatrix}
 \end{equation}  

The y and z components of the dipole moment vanish in Eq. (\ref{eq:10}) because we approximate the NS by a point mass moving the x-direction. To calculate the polarizations of gravitational radiation, we need to move from the inertial frame to the wave frame. This is done by

\begin{eqnarray} 
\label{eq:11}
Q_W^{ij} (t') &=& S \cdot Q^{ij}(t') \cdot S^T
\nonumber \\
D_W(t') &=& S \cdot D (t')
\end{eqnarray} 

where $Q_W^{ij}(t')$ is the quadrupole moment tensor in the wave frame, $D_W(t')$ is the dipole moment in the wave frame, $S$ is the rotation matrix given by

    \begin{equation} 
 \label{eq:12}
S  =  \begin{bmatrix} \cos \iota & 0 & - \sin \iota  \\ 0 & 1  & 0 \\ \sin \iota & 0 & \cos \iota  \end{bmatrix}
 \end{equation} 

and $S^T$ is the transpose of $S$. $\iota$ is the angle between wave and z-axis of the inertial frame.

The polarizations in BD theory are given by \cite{Verma}

\begin{eqnarray} 
\label{eq:13}
h_{+} (t) &=& \frac{G}{r c^{4}} (1 - \zeta) 
(\ddot{Q}^{xx}_{W} (t^{\prime}) - \ddot{Q}^{yy}_{W} (t^{\prime})) 
\nonumber \\
h_{\times} (t) &=& \frac{2 G}{r c^{4}} (1 - \zeta)  \ddot{Q}^{xy}_{W} (t^{\prime}) \nonumber \\
h_{S} (t) &=& \frac{2G}{r c^{2}}  \zeta \left[  M(t^{\prime})  +  \frac{1}{c} \dot{D}_{W}^{z} (t^{\prime})  - \frac{1}{2 c^2}  \ddot{Q}^{zz}_{W}  (t^{\prime}) \right]
\end{eqnarray} 

where $h_{+} (t)$ is the plus polarization, $h_{\times} (t)$ is the cross polarization, $h_{S} (t)$ is the scalar polarization and $r$ is the distance of the source from the detector. In the above equations, $t$ is the time when polarizations are measured and $t'$ represents the retarded time. After some algebraic manipulations and ignoring the mass monopole term, we obtain

\begin{eqnarray} 
\label{eq:14}
h_{+} (t) &=& \frac{G}{r c^{4}} (1 - \zeta) m  \left[ 2 u^2 + 3 a_t^2 t'^2 + 6 u a_t t'  \right] \cos^2 \iota \nonumber \\
h_{\times} (t) &=& 0 \nonumber \\
h_{S} (t) &=& \frac{2G}{r c^{2}}  \zeta \left[ \frac{m}{c} (u + a_t t') \sin \iota  - \frac{m}{c^2} \left( 3 \sin^2 \iota -1  \right)  \left ( \frac{1}{3} u^2 + \frac{1}{2} a_t^2 t'^2 + u a_t t'  \right) \right]
\end{eqnarray}

\section{Analysis}

To constrain the BD parameter, we use only the expression of scalar polarization. On the other hand, the quadrupole term in the scalar polarizations is much weaker than the dipole term  due to extra powers of c in the denominator. Hence, ignoring the quadrupole contribution, we obtain

    \begin{equation} 
 \label{eq:15}
h_{S} (t) \approx \frac{2G}{r c^{2}}  \zeta \left[ \frac{m}{c} (u + a_t t) \sin \iota \right]
 \end{equation} 

The power emitted $P^{(S)}$ in the scalar radiation is given by \cite{Will-1977}, \cite{Verma-rod-BD}

    \begin{equation} 
 \label{eq:16}
 \frac{dP^{(S)}}{dA} = \frac{c^3}{16 \pi \zeta G  } <   \dot{h}^2_{s}(t) >
 \end{equation} 

where $dA = r^2 \int _{\rho = 0}^{2 \pi} \int _{\iota = - \frac{\pi}{2}}^{  \frac{\pi}{2}}  \cos \iota d \iota d \rho  $ and $ < \cdot > $ implies the time average.

By substituting the first derivative square of Eq. (\ref{eq:15}) in Eq. (\ref{eq:16}) and performing the integration, we get

    \begin{equation} 
 \label{eq:17}
 P^{(S)} = \frac{1}{3} \zeta \frac{G}{c^3} m^2 a_t^2
 \end{equation} 
 
 Using the fact that power emitted is the rate of change of kinetic energy, we write
 
     \begin{equation} 
 \label{eq:18}
\frac{d}{dt} \left( \frac{1}{2} m v^2  \right) = - \frac{1}{3} \zeta \frac{G}{c^3} m^2 a_t^2
 \end{equation} 
 
 where minus sign on the right hand side of Eq. (\ref{eq:18}) implies that radiation takes away energy from the system. Using the fact that $\frac{dv}{dt} = v \frac{dv}{dx}$, we get 
 
\begin{equation} 
 \label{eq:19}
\int_{u}^{V} v^2 dv = - \int_{0}^{x} \frac{1}{3} \zeta \frac{G}{c^3} m a_t^2 dx
 \end{equation} 
 
 where $u$ is the initial velocity (moment we start making observation), $V$ is the moment we stop making observation and $x$ represents the distance traveled by the NS during the observation starting from the origin of the inertial frame. A simple integration yields

 \begin{equation} 
 \label{eq:20}
V = \left[ u^3 - \zeta \frac{G m a_t^2}{c^3} x \right]^{\frac{1}{3}}
 \end{equation}

 The velocity $V$ in Eq. (\ref{eq:20}) is the final theoretical value. But in reality, the observed final velocity $V_{obs}$ along the path should be less than the theoretical value for the following reason. We obtain the theoretical velocity $V$ by considering only the scalar wave. But LIGO and Virgo detectors have already confirmed the existence of tensor polarizations. So, the power carried away by the tensor polarizations further decreases the velocity along the path and we get
 
  \begin{equation} 
 \label{eq:21}
V_{obs} < V
 \end{equation} 

or,

  \begin{equation} 
 \label{eq:22}
V_{obs} < \left[ u^3 - \zeta \frac{G m a_t^2}{c^3} x \right]^{\frac{1}{3}}
 \end{equation} 
 
By rearranging the terms in the above inequality, we obtain

  \begin{equation} 
 \label{eq:23}
\zeta < \frac{c^3}{G} \frac{\left( u^3 - V_{obs}^3  \right)}{ m a_t^2 x }  
 \end{equation}

 Since $a_t$ can be considered as a constant for a short distance, we can write it as

   \begin{equation} 
 \label{eq:24}
a_t = \frac{V_{obs} - u}{t} 
 \end{equation} 
 
By substituting Eq. (\ref{eq:24}) in Eq. (\ref{eq:23}), we get
 
  \begin{equation} 
 \label{eq:25}
\boxed{  \zeta < \frac{c^3}{G} \frac{\left( u^3 - V_{obs}^3  \right)}{ \left( u - V_{obs}    \right)^2 }  \frac{t^2}{mx}}
 \end{equation}

  In the Eq. (\ref{eq:25}), the parameters $u$ is the initial observational velocity when we start making the observation, $V_{obs}$ is the final observed velocity (when we stop making the observation), $t$ is the time duration of the observation, $m$ is the mass of the NS orbiting around the SMBH and $x$ is the distance traveled by the NS perpendicular to the line of sight. For a short distance along the orbit, we can write $x \approx R \theta$ using the small angle approximation. Here, R is the radius of the orbit and $\theta$ is the angle subtended by the short parth from the center of the SMBH. The above calculations give a constraint on the BD parameter using a compact object revolving around a SMBH whereas  the constraint on the BD parameter from the Cassini mission is $ \zeta < 0.0000125$ which corresponds to testing in the theory in the Solar system.
 
\section*{Conclusions} 

We study the case when a NS revolved around a galactic central engine. We start by using the theory of gravitational radiation and express the BD parameter in terms of parameters which can be observed through astronomical observations. This formula can be used to put a constraint on the BD parameter in the radiation regime. Although we started with the theory of gravitational radiation, we don't need any gravitational wave detection to impose this constraint. The dynamics of a NS star around a SMBH can be observed by astronomical telescopes.

\section*{Acknowledgement}

I cordially thank Prof. Andrzej Królak for his valuable guidance during my PhD studies and for providing me with enough room and motivation for independent thinking. I express my deep gratitude to the LIGO-Virgo-Kagra collaboration which has offered me a precious opportunity to work with eminent physicists around the world. This work was supported by the Polish National Science Centre Grant No. 2017/26/M/ ST9/0097.





\begin{thebibliography}{999}



\bibitem{Brans-Dicke} Brans, C.; Dicke, R.H.  Mach's Principle and a Relativistic Theory of Gravitation. {\em Phys. Rev. Lett.} {\bf 1961}, {\em 124}, 925



\bibitem{Jordan} Jordan, P. Zum gegenw\"artigen Stand der Diracschen kosmologischen Hypothesen. {\em Z. Phys} {\bf 1959}, {\em 157}, 112–121.

\bibitem{Fierz} Fierz, M.  \"Uber die physikalische Deutung der erweiterten Gravitationstheorie P.
Jordans. {\em Helv. Phys. Acta} {\bf 1956}, {\em 29}, 128–134


\bibitem{Isi} Isi, M.; Pitkin, M.; Weinstein, A.J.  Probing Dynamical Gravity with the Polarization of Continuous Gravitational Waves. {\em Phys. Rev. D.} {\bf 2017}, {\em 96}, 042001.



\bibitem{Bertotti-Tartora} Bertotti, B.; Iess, L.; Tortora, P. A test of general relativity using radio links with the Cassini spacecraft. {\em Nature} {\bf 2003}, {\em 425}, 374–376.


\bibitem{GR test GWTC3} Abbott, R.; Abe, H.; Acernese, F.; Ackley, K.; Adhikari, N.; Adhikari, R.X.; Adkins, V.K.; Adya, V.B.; Affeldt, C.; Agarwal, D.; et al. Tests of General Relativity with GWTC-3. arXiv:2112.06861


\bibitem{GR-test-GW170817} Abbott, B.P; Abbott, R.; Abbott, T.D; Acernese, F.; Ackley, K.; Adams, C.; Adams, T.; Addesso, P.; Adhikari, R.X.; Adya, V.B.; et al. Tests of General Relativity with GW170817. {\em Phys. Rev. Lett.} {\bf 2019}, {\em 123}, 011102.

\bibitem{Bertolami} Bertolami, O.; Martins, P.J. 
Nonminimal coupling and quintessence {\em Phys. Rev. D} {\bf 2000}, {\em 61}, 064007.

\bibitem{JBD-cosmo} Almeida, C.R.; Galkina, O.; Fabris, J.C. Quantum and Classical Cosmology in the Brans–Dicke Theory. {\em Universe}, {\bf 2021}, {\em 7}, 286.

\bibitem{Sola} Sola, J.; Gomez-Valent, A.; Perez, J.D.C.; Moreno-Pulido, C. Brans-Dicke cosmology with a $\Lambda$ - term: a possible solution to $ \Lambda $CDM tensions. {\em Class. Quantum Grav}, {\bf 2020}, {\em 37}, 245003.

\bibitem{BD-LVK} Abbott, R.; Abe, H.; Acernese, K.; Ackley, N.; Adhikari, N.; Adhikari, R.X.; Adkins, V.K.; Adya, V.B.; Affeldt, C.; Agarwal, D.; et al. Searches for Gravitational Waves from Known Pulsars at Two Harmonics in the Second and Third LIGO-Virgo Observing Runs. arXiv:2111.13106



\bibitem{Wex} Wex, N. Testing Relativistic Gravity with Radio Pulsars. arXiv:1402.5594   



\bibitem{Verma} Verma, P. Probing Gravitational Waves from Pulsars in Brans-Dicke Theory. {\em Universe} {\bf 2021}, {\em 7(7)}, 235


\bibitem{Will-1977} Will, C.M. Gravitational radiation from binary systems in alternative metric theories of gravity: dipole radiation and the binary pulsar. {\em ApJ.} {\bf 1977}, {\em 214}, 826-839. 

\bibitem{Verma-rod-BD} Verma, P. A swinging rod in Brans-Dicke theory. {\em Annalen Der Physik} {\bf 2022}, 2100600.  



\bibitem{Krolak} Kr\'olak, A; Jaranowski, P. {\em Analysis of Gravitational-Wave Data}, Cambridge University Press, Cambridge, UK, 2009. 



\bibitem{Will2014} Will, C.M. The Confrontation between General Relativity and Experiment. {\em Living Rev. Relativ.} {\bf 2014}, {\em 17}, 4

\bibitem{Poisson}Poisson, E and Will, C.M. {\em Gravity Newtonian, Post-Newtonian, Relativistic}; Cambridge University Press, Cambridge, UK, 2014 

\bibitem{Misner}Misner, C.W.; Thorne, K.S.; Wheeler, J.A.  {\em Gravitation}; W. H. Freeman and Company, San Francisco, USA, 1973 



\bibitem{Fuji-Maeda} Fuji, Y.; Maeda K. {\em The scalar-tensor theory of gravitation }, Cambridge University Press, Cambridge, UK, 2004.   





\end{thebibliography}
\end{document}